\documentclass{sig-alternate}
\newcommand{\ignore}[1]{}
\usepackage[pass]{geometry}
\usepackage{fancyhdr}
\usepackage[normalem]{ulem}
\usepackage[hyphens]{url}
\usepackage{hyperref}

\usepackage{graphicx}
\usepackage{subfig}
\usepackage{authblk}
\graphicspath{{./figures/}}
\usepackage{url}
\usepackage{colortbl}
\definecolor{mygray}{gray}{.88}
\usepackage{multirow}

\newcommand{\tabincell}[2]{\begin{tabular}{@{}#1@{}}#2\end{tabular}}
\newcommand{\comp}{\hspace{0pt}\nolinebreak-\hspace{0pt}}

\title{A Dwarf-based Scalable Big Data Benchmarking Methodology}
\author[1,2]{\normalsize Wanling Gao}
\author[1,2]{Lei Wang}
\author[1,2]{Jianfeng Zhan \thanks{The corresponding author is Jianfeng
Zhan.}}
\author[1,2]{Chunjie Luo}
\author[1,2]{Daoyi Zheng}
\author[4]{Zhen Jia}
\author[1,2]{Biwei Xie}
\author[1,2]{Chen Zheng}
\author[5,6]{Qiang Yang}
\author[3]{Haibin Wang}
\affil[1]{State Key Laboratory of Computer Architecture (Institute of
Computing Technology, Chinese Academy of Sciences)
\authorcr \{gaowanling, wanglei\_2011, zhanjianfeng, luochunjie, zhengdaoyi, xiebiwei, zhengchen\}@ict.ac.cn}
\affil[2]{University of Chinese Academy of Sciences}
\affil[3]{Huawei, benjamin.wanghaibin@huawei.com}
\affil[4]{Princeton University, zhenj@princeton.edu}
\affil[5]{Beijing Academy of Frontier Science \& Technology, yangqiang@mail.bafst.com}
\affil[6]{Chinese Industry, Intelligence and Information Data Technology Corporation}

\begin{document}
\maketitle
\pagestyle{plain}

\begin{abstract}
Different from the traditional benchmarking methodology that creates a new benchmark or proxy for every possible workload, this paper presents a scalable big data benchmarking methodology.
Among a wide variety of big data analytics workloads, we identify eight big data dwarfs, each of which captures the common requirements of each class of
unit of computation while being reasonably divorced from individual implementations.
%
We implement the eight dwarfs on different software stacks, e.g., OpenMP, MPI, Hadoop as the dwarf components. For the purpose of architecture simulation, we construct and tune big data proxy benchmarks   using the directed acyclic graph (DAG)-like combinations of the  dwarf components with different weights to mimic the benchmarks in BigDataBench. Our proxy benchmarks preserve the micro-architecture, memory, and I/O characteristics, and they
shorten the simulation time by 100s
times while maintain the average micro-architectural
data accuracy above 90 percentage on both X86\_64 and
ARMv8 processors. We will open-source the big data dwarf components
and proxy benchmarks soon.

\end{abstract}

\section{Introduction}

\begin{figure}[!t]
\centering
\includegraphics[scale=0.56]{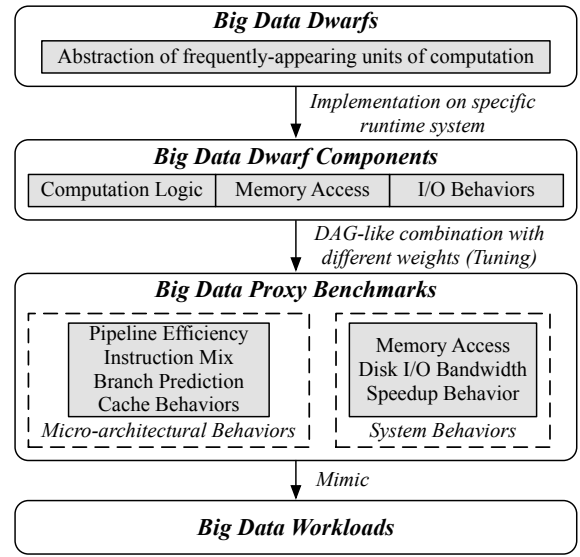}
\caption{Dwarf-based Scalable Big Data Benchmarking Methodology.} 
\label{overall}
\end{figure}

Benchmarking is the foundation of system and architecture evaluation. The
benchmark is run on several different systems, and the performance and price of each system is
measured and recorded~\cite{gray1993database}. The most simplest benchmarking methodology is to  creates a new benchmark for every possible workload.
PARSEC~\cite{bienia11benchmarking},  BigDataBench~\cite{wang2014bigdatabench} and CloudSuite~\cite{ferdman2011clearing} are put together in this way. This methodology has two drawbacks.  First, the modern real-world workloads change very frequently, and we need keep expanding the benchmark set with emerging workloads. So it is very difficult to build a comprehensive benchmark suite.  Second, whatever early in the design process or later in the system evaluation, it is  time-consuming to run a comprehensive benchmark suite. The complex software stacks of the modern workloads aggravate this issue. For example, running  big data  benchmarks on the simulators needs prohibitively long execution.   We  use \emph{benchmarking scalability}~\footnote{Here, the meaning of scalable differs from scaleable.  As one of four properties of domain-specific benchmarks defined by Jim Gray~\cite{gray1993database}, the latter refers to  scaling  the benchmark up to larger systems}---which can be measured in terms of the cost of building and running a comprehensive benchmark suite---to denote this challenge.

For scalability, proxy benchmarks, abstracted from real workloads, are widely adopted. Kernel benchmarks---small programs extracted from large application programs~\cite{lilja2005measuring,hennessy2011computer}---are widely used in high performance
computing. However, it is insufficient to completely reflect workload behaviors and has limited usefulness in making overall comparisons~\cite{bailey1991parallel,lilja2005measuring}, especially for complex big data workloads.
Another attempt is to generate
synthetic traces or benchmarks based on the workload trace to mimic workload behaviors, and they propose the methods like statistic simulation~\cite{conte1996reducing,wunderlich2003smarts,yu2009tss,sherwood2002automatically} or sampling~\cite{skadron2003challenges,eeckhout2004control,eeckhout2000performance} to accelerate this trace.  It omits the computation logic of the original workloads, and has several disadvantages.  First, trace-based synthetic benchmarks target
 specific architectures, and different architecture configuration will generate different synthetics benchmarks. Second, synthetic benchmark cannot be deployed across over different architectures, which makes it hard for cross-architecture comparisons. Third, workload trace has strong dependency on data input, which make it hard to mimic real benchmark with diverse data inputs. 

Too complex workloads are not helpful for both reproducibility and interpretability of performance data. So identifying abstractions of frequently-appearing units of computation, which we call big data dwarfs,
is an important step toward fully understanding big data analytics workloads.
The concept of dwarf, which is firstly proposed by Phil Colella~\cite{colella2004defining}, is thought to be not only a highly abstraction of workload patterns, but also a minimum set of necessary functionality~\cite{minimumSet}. 
A dwarf captures the common requirements of each class
of unit of computation while being reasonably divorced
from individual implementations~\cite{asanovic2006landscape}~\footnote{Our definition has a subtle difference from the definition in the Berkeley report~\cite{asanovic2006landscape}}. 
For example, transformation from one domain to another domain contains a class of unit of computation, such as fast fourier transform (FFT) and discrete cosine transform (DCT).
Intuitively, we can build big data proxy benchmarks
 on the basis of big data dwarfs.  Considering complexity, diversity and fast-changing characteristics of big data analytics workloads, we propose a dwarf-based scalable big data benchmarking methodology, as shown in Fig.~\ref{overall}.

After thoroughly analyzing a majority of workloads in five typical big data application domains (search engine, social network, e-commerce, multimedia and bioinformatics), we identify eight  big data dwarfs, including \emph{matrix}, \emph{sampling}, \emph{logic}, \emph{transform}, \emph{set}, \emph{graph}, \emph{sort} and \emph{basic statistic} computations that frequently appear~\footnote{We acknowledge our eight dwarfs may be not enough for other applications we fail to investigate. We will keep on expanding this dwarf set.}, the combinations of which describe most of big data workloads we investigated.
We implement eight dwarfs on different software stacks like MPI, Hadoop with diverse data generation tools for text, matrix and graph data, which we call the dwarf components. As software stack has great influences on workload behaviors~\cite{wang2014bigdatabench,jia_bigDataBench_subset}, our dwarf  component implementation considers the execution model of software stacks and the programming styles of workloads.
 For feasible simulation, we choose the OpenMP version of dwarf components to construct proxy benchmarks as OpenMP is  much more light-weight than the other programming frameworks in terms of binary size and also supported by a majority of simulators~\cite{abellan2016electro}, such as GEM5~\cite{binkert2011gem5}.
As  a node represents original data set or an intermediate data set being processed and a edge represents the dwarf components, we use the DAG-like combination of one or more dwarf components with different weights to form our proxy benchmarks.
We provide an auto-tuning tool to generate qualified proxy benchmarks satisfying the simulation time and micro-architectural data accuracy requirements, through an iteratively execution including the adjusting and feedback stages.

On the typical X86\_64 and ARMv8 processors, the proxy benchmarks have about 100s times runtime speedup with the average micro-architectural data accuracy above 90\% with respect to the benchmarks from BigDataBench.
Our proxy benchmarks have been applied to the ARM processor design and implementation in our industry partnership.
 Our contributions are two-fold as follows:

\begin{itemize}
\item We propose a dwarf-based scalable big data benchmarking methodology, using a DAG-like combination of the dwarf components with different weights to mimic big data analytics workloads.
\item We identify eight  big data dwarfs, and implement the dwarf components for each dwarf on different software stacks with diverse data inputs. Finally we construct the  big data proxy benchmarks, which reduces simulation time and guarantees performance data accuracy for micro-architectural simulation.
\end{itemize}

The rest of the paper is organized as follows.
Section 2 presents the methodology. Section 3 performs evaluations on a five-node X86\_64 cluster. In Section 4, we report using the proxy benchmarks on the ARMv8 processor.
Section 5 introduces the related work. Finally, we draw a conclusion in Section 6.

\section{Benchmarking Methodology}

\subsection{Methodology Overview} \label{overview}

\begin{figure*}[!t]
\centering
\includegraphics*[scale=1]{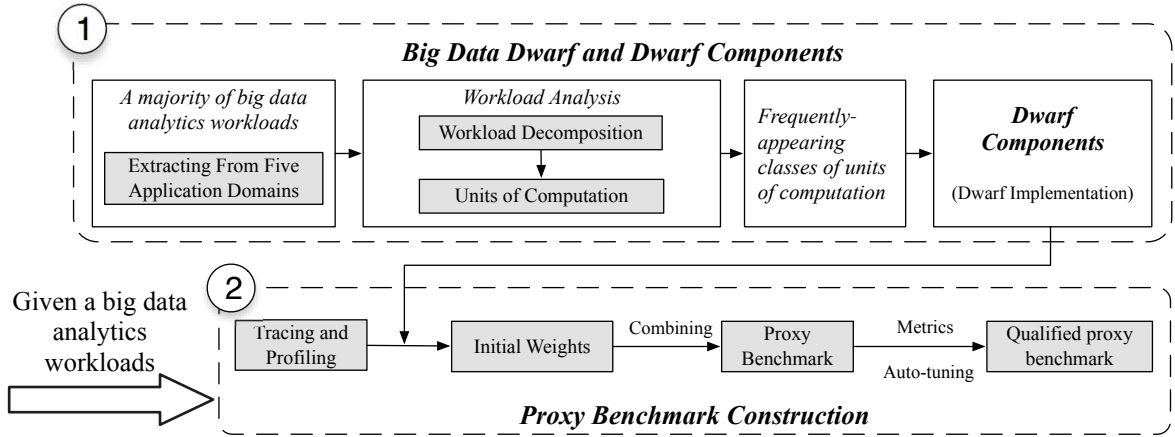}
\caption{Methodology Overview.} 
\label{dwarfidentify}
\end{figure*}

In this section, we illustrate our dwarf-based scalable big data benchmarking methodology, including dwarf identification, dwarf component implementation, and proxy benchmark construction.
Fig.~\ref{dwarfidentify} presents our benchmarking methodology.

First, we analyze a majority of big data analytics workloads through workload decomposition and conclude frequently\comp appearing classes of units of computation, as big data dwarfs. 
Second, we provide the dwarf implementations using different software stacks, as the dwarf components. 
Finally, we construct proxy benchmarks using the DAG-like combinations of the dwarf components with different weights to mimic the real-world workloads.
A DAG-like structure uses a node to represent original data set or intermediate data set being processed, and uses a edge to represent the dwarf components.

Given a big data analytics workload, we get the running trace and execution time through the tracing and profiling tools.
According to the execution ratios, we identify the hotspot functions and correlate them to the code fragments of the workload through bottom-up analysis. Then we analyze these code fragments to choose the specific dwarf components and set their initial weights according to execution ratios.
Based on the dwarf components and initial weights, we construct our proxy benchmarks using a DAG-like combination of dwarf components.
For the targeted performance data accuracy, such as cache behaviors or I/O behaviors, we provide an auto-tuning tool to tune the parameters of the proxy benchmark, and generate qualified proxy benchmark to satisfy the requirements.

\begin{table*}[htb]
\caption{Eight Classes of Units of Computation.}\label{statistical}
\renewcommand\arraystretch{1.4}
\footnotesize
\center
\begin{tabular}{|p{1.12in}|p{1.4in}|p{2.1in}|p{1.7in}|}
  \hline
  \textbf{Catergory}   & \textbf{ Application Domain } &  \textbf{Workload } &  \textbf{ Unit of Computation } \\
  \hline
  \multirow{2}{*}{\tabincell{l}{Graph Mining}}  &  \multirow{2}{*}{\tabincell{l}{ Search Engine \\ Community Detection }}  & PageRank  & Matrix, Graph, Sort \\  \cline{3-4}
                                                &   & BFS, Connected component(CC)      &  Graph \\ \cline{3-4}
  \hline

  \multirow{2}{*}{\tabincell{l}{Deminsion Reduction}}  &  \multirow{2}{*}{\tabincell{l}{ Image Processing \\ Text Processing }}  & Principal components analysis(PCA)  & Matrix \\  \cline{3-4}
                                                &   & Latent dirichlet allocation(LDA)  &  Basic Statistic, Sampling \\ \cline{3-4}
  \hline

  \multirow{2}{*}{\tabincell{l}{Deep Learning}} &  \multirow{2}{*}{\tabincell{l}{ Image Recognition \\ Speech Recognition }} & Convolutional neural network(CNN)  &  Matrix, Sampling, Transform \\  \cline{3-4}
                                                &  & Deep belief network(DBN)         & Matrix, Sampling \\ \cline{3-4}
  \hline

  \multirow{3}{*}{\tabincell{l}{Recommendation}} & \multirow{3}{*}{\tabincell{l}{Association Rules Mining \\ Electronic Commerce }}  & Aporiori  &  Basic Statistic, Set \\  \cline{3-4}
                                                             &  & FP-Growth &  Graph, Set, Basic Statistic \\ \cline{3-4}
                                                             &  & Collaborative filtering(CF) &  Graph, Matrix \\ \cline{3-4}
  \hline

  \multirow{4}{*}{\tabincell{l}{Classification}} &  \multirow{4}{*}{\tabincell{l}{ Image Recognition \\ Speech Recognition \\ Text Recognition }} & Support vector machine(SVM)  &  Matrix \\  \cline{3-4}
                                                 &  & K-nearest neighbors(KNN) &  Matrix, Sort, Basic Staticstic \\ \cline{3-4}
                                                 &  & Naive bayes  &  Basic Statistic \\ \cline{3-4}
                                                 &  & Random forest  &  Graph, Basic Statistic  \\ \cline{3-4}
                                                 &  & Decision tree(C4.5/CART/ID3)  &  Graph, Basic Statistic  \\ \cline{3-4}
  \hline

  \multirow{1}{*}{\tabincell{l}{Clustering}} & \multirow{1}{*}{\tabincell{l}{ Data Mining }} &  K-means  & Matrix, Sort  \\ \cline{3-4}
  \hline

  \multirow{4}{*}{\tabincell{l}{Feature Preprocess}} & \multirow{4}{*}{\tabincell{l}{ Image Processing \\ Signal Processing \\ Text Processing}}  & Image segmentation(GrabCut)  & Matrix, Graph \\  \cline{3-4}
                                                     &  & Scale-invariant feature transform(SIFT)  & Matrix, Transform, Sampling, Sort, Basic Statistic \\  \cline{3-4}
                                                     &  & Image Transform  & Matrix, Transform \\  \cline{3-4}
                                                     &  & Term Frequency-inverse document frequency (TF-IDF)   & Basic Statistic \\  \cline{3-4}
  \hline

  \multirow{2}{*}{\tabincell{l}{Sequence Tagging}} &\multirow{2}{*}{\tabincell{l}{ Bioinformatics \\ Language Processing }}  & Hidden Markov Model(HMM)  &  Matrix \\  \cline{3-4}
                                               &  & Conditional random fields(CRF) & Matrix, Sampling \\ \cline{3-4}
  \hline


  Indexing & Search Engine & Inverted index, Forward index & Basic Statistic, Logic, Set, Sort\\
  \hline

  \multirow{4}{*}{\tabincell{l}{Encoding/Decoding}} & \multirow{4}{*}{\tabincell{l}{ Multimedia Processing \\ Security \\ Cryptography \\ Digital Signature }}  & MPEG-2  & Matrix, Transform \\  \cline{3-4}
                                                    &   & Encryption   & Matrix, Logic \\  \cline{3-4}
                                                    &   & SimHash, MinHash  & Set, Logic \\  \cline{3-4}
                                                    &  & Locality-sensitive hashing(LSH)  & Set, Logic \\ \cline{3-4}
  \hline

  \multirow{1}{*}{\tabincell{l}{Data Warehouse}} & \multirow{1}{*}{\tabincell{l}{  Business intelligence }}  &  Project, Filter, OrderBy, Union  &  Set, Sort \\  \cline{3-4}
  \hline


\end{tabular}
\end{table*}

\subsection{Big Data Dwarf and Dwarf Components}

In this section, we illustrate the big data dwarfs and corresponding dwarf component implementations.

After singling out a broad spectrum of big data analytics workloads (machine learning, data mining, computer vision and natural language processing) through investigating five application domains (search engine, social network, e-commerce, multimedia, and bioinformatics), we analyze these workloads and decompose them to multiple classes of units of computation.
According to their frequency and importance, we finalize eight big data dwarfs, which are abstractions of frequently\comp appearing classes of units of computation.
Table~\ref{statistical} shows the importance of eight classes of units of computation (dwarfs) in a majority of big data analytics workloads.
We can find that these eight dwarfs are major classes of units of computations in a variety of big data analytics workloads.

\subsubsection{Eight Big Data Dwarfs} \label{bigdatadwarf}

In this subsection, we summarize eight big data dwarfs frequently appearing in big data analytics workloads.

\textbf{Matrix Computations} In big data analytics, many problems involve matrix computations, such as matrix multiplication and matrix transposition.

\textbf{Sampling Computations} Sampling plays an essential role in big data processing, which can obtain an approximate solution when one problem cannot be solved by using analytical method.

\textbf{Logic Computations} We name computations performing bit manipulation as logic computations, such as hash, data compression and encryption. 

\textbf{Transform Computations} The transform computations here mean the conversion from the original domain (such as time) to another domain (such as frequency). Common transform computations include discrete fourier transform (DFT), discrete cosine transform (DCT) and wavelet transform.

\textbf{Set Computations} In mathematics, set means a collection of distinct objects. Likewise, the concept of set is also widely used in  computer science. 
For example, similarity analysis of two data sets involves set computations, such as Jaccard similarity. Furthermore, fuzzy set and rough set play very important roles in computer science. 

\textbf{Graph Computations} A lot of applications involve graphs,
with nodes representing entities and edges representing dependencies.
Graph computations are notorious for having irregular memory access patterns.

\textbf{Sort Computations} Sort is widely used in many areas. Jim Gray thought sort is the core of modern databases~\cite{asanovic2006landscape}, which shows its fundamentality. 

\textbf{Basic Statistic Computations} Basic statistic computations are used to obtain the summary information through statistical computations, such as counting and probability statistics.

\subsubsection{Dwarf Components} \label{component}

\begin{figure}[!t]
\centering
\includegraphics*[scale=0.61]{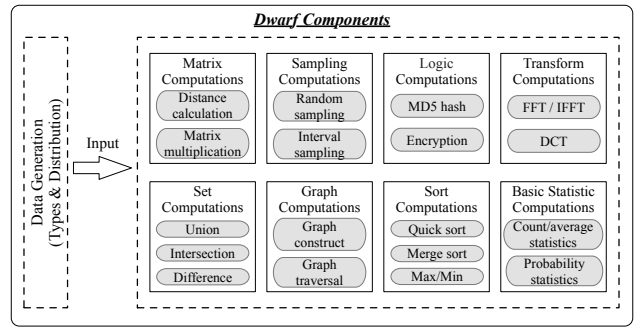}
\caption{The Overview of the Dwarf Components.} 
\label{library}
\end{figure}

Fig.~\ref{library} presents the overview of our dwarf components, which consist of two parts-- data generation tools and dwarf implementations. The data generation tools provide various data inputs with different data types and distributions to the dwarf components, covering text, graph and matrix data. 
Since software stack has great influences on workload behaviors~\cite{wang2014bigdatabench,jia_bigDataBench_subset}, our dwarf component implementation considers the execution model of software stacks and the programming styles of workloads using specific software stacks.
Fig.~\ref{library} lists all  dwarf components. For example, we provide distance calculation (i.e. euclidian, cosine) and matrix multiplication for matrix computations.
For simulation-based architecture research, we provide light-weight implementation of dwarf components using the OpenMP~\cite{dagum1998openmp} framework as OpenMP has several advantages. First, it is widely used. Second, it is much more light-weight than the other programming frameworks in terms of binary size. Third, it is supported by a majority of simulators~\cite{abellan2016electro}, such as GEM5~\cite{binkert2011gem5}.
For the purpose of system evaluation, we also implemented the dwarf components on several other software stacks including MPI~\cite{gropp1996high}, Hadoop~\cite{hadoopweb} and Spark~\cite{spark}.

\subsection{Proxy Benchmarks Construction}\label{tuning}

\begin{figure*}[!t]
\centering
\includegraphics*[scale=0.77]{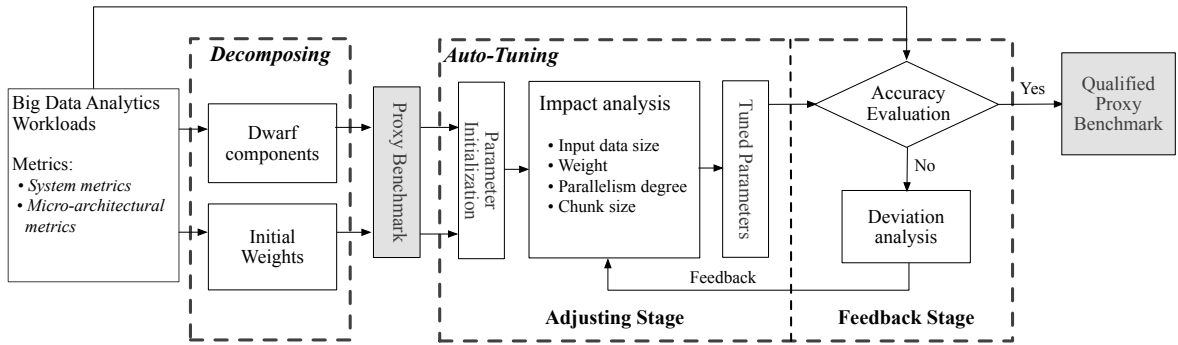}
\caption{Proxy Benchmarks Construction.} 
\label{tuning}
\end{figure*}

Fig.~\ref{tuning} presents the process of proxy benchmark construction, including \emph{decomposing} process and \emph{auto-tuning} process.
Given a big data analytics workload, we obtain its hotspot functions and execution time through a multi-dimensional tracing and profiling method, including runtime tracing (e.g. JVM tracing and logging), system profiling (e.g. CPU time breakdown) and hardware profiling (e.g. CPU cycle breakdown).
Based on the hotspot analysis, we correlate the hotspot functions to the code fragments of the workload and choose the corresponding dwarf components through analyzing the computation logic of the code fragments.
Our proxy benchmark is a DAG-like combination of the selected dwarf components with initial weights setting by their execution ratios.

We measure the proxy benchmark's accuracy by comparing the performance data of the proxy benchmark with those of the original workloads in both system and micro-architecture level.
In our methodology, we can choose different performance metrics  to measure the accuracy according to the concerns about different behaviors of the workload. For example, if our proxy benchmarks are to focus on cache behaviors of the workload, we can choose the  metrics that reflect cache behaviors like cache hit rate  to tune a qualified proxy benchmark.
To tune the  accuracy---making  it more similar to the original workload,  we further provide an auto-tuning tool with four parameters to be tuned (listed in Table \ref{factor}). We found those four parameters play essential roles in the system and architectural behaviours of a certain workload.
The tuning process is an iteratively execution process which can be separated into three stages: parameter initialization, adjusting stage and feedback stage. We elaborate them in detail as below:

\textbf{Parameter Initialization} We initialize the four parameters(\emph{Input Data Size}, \emph{Chunk Size}, \emph{Parallelism Degree} and \emph{Weight}) according to the configuration of the original workload.
We scale down the input data set and chunk size of the original workloads to initialize \emph{Input Data Size} and \emph{Chunk Size}.
The \emph{Parallelism Degree} is initialized as the parallelism degree of the original workload.
To guarantee the rationality of the weights of each dwarf component, we initialize the \emph{weights} proportional to their corresponding execution ratios. For example, in Hadoop TeraSort, the weight is  70\% of sort computation, 10\% of sampling computation, and 20\% of  graph computation, respectively.
Note that the \emph{weight} can be adjusted within a reasonable range (e.g. plus or minus 10\%) during the tuning process.

\begin{table}[htb]
\caption{Tunable Parameters for Each Dwarf Component.}\label{factor}
\renewcommand\arraystretch{1.2}
\center
\footnotesize
\begin{tabular}{|p{1.1in}|p{1.8in}|}
  \hline
  \textbf{Parameter} & \textbf{Description} \\
  \hline
  Input Data Size & The input data size for each dwarf component \\
  \hline
  Chunk Size & The data block size processed by each thread for each dwarf component \\
  \hline
  Parallelism Degree & The process and thread numbers for each dwarf component \\
  \hline
  Weight & The contribution for each dwarf component \\
  \hline
\end{tabular}
\end{table}

\textbf{Adjusting Stage}

We introduce a decision tree based mechanism to assist the auto-tuning process.
The tool learns the impact of each parameter on all metrics and build a decision tree through \emph{Impact analysis}.
The learning process changes one parameter each time and execute multiple times to characterize the parameter's impact on each metric.
Based on the impact analysis, the tool builds a decision tree to determine which parameter to tune if one metric has large deviation.
After that, the \emph{Feedback Stage} is activated to evaluate the proxy benchmark with tuned parameters. If it does not satisfy the requirements, the tool will adjust the parameters to improve the accuracy using the decision tree in the adjusting stage.

\textbf{Feedback Stage}
In the feedback stage, the tool evaluates the accuracy of the current proxy benchmark with specific parameters.
If the deviations of all metrics are within the setting range (e.g. 15\%), the auto-tuning process is finished.
Otherwise, the metrics with large deviations will be fed back to the adjusting stage.
The adjusting and feedback process will iterate until reaching the specified accuracy, and the finalized proxy benchmark with the final parameter settings is our qualified proxy benchmark.

\subsection{Proxy Benchmarks Implementation}

\begin{table*}[htbp]
\caption{Four Hadoop Benchmarks from BigDataBench and Their Corresponding Proxy Benchmarks.}\label{data_workloads}
\renewcommand\arraystretch{1.2}
\center
\footnotesize
\begin{tabular}{|p{0.61in}|p{1in}|p{0.6in}|p{1.41in}|p{2.52in}|}
  \hline
  \tabincell{l}{Big Data\\Benchmark} & \tabincell{l}{Workload Patterns} & \tabincell{l}{Data Set} & \tabincell{l}{Involved Dwarfs} & Involved Dwarf Components  \\
  \hline
  \tabincell{l}{Hadoop\\TeraSort} &\tabincell{l}{I/O Intensive} & \tabincell{l}{Text} & \tabincell{l}{Sort computations\\Sampling computations\\Graph computations} & \tabincell{l}{Quick sort; Merge sort\\Random sampling; Interval sampling\\Graph construction; Graph traversal} \\
  \cline{1-5}
  \tabincell{l}{Hadoop\\Kmeans} &\tabincell{l}{CPU Intensive} & \tabincell{l}{Vectors} & \tabincell{l}{Matrix computations\\Sort computations\\Basic Statistic} & \tabincell{l}{Vector euclidean distance; Cosine distance\\Quick sort; Merge sort\\Cluster count; Average computation}  \\
  \cline{1-5}
  \tabincell{l}{Hadoop\\PageRank}&\tabincell{l}{Hybrid} &\tabincell{l}{Graph} & \tabincell{l}{Matrix computations\\Sort computations\\Basic Statistic } & \tabincell{l}{Matrix construction; Matrix multiplication\\Quick sort; Min/max calculation\\Out degree and in degree count of nodes} \\
  \cline{1-5}
  \tabincell{l}{Hadoop\\SIFT} &\tabincell{l}{CPU Intensive\\Memory Intensive}&\tabincell{l}{Image} & \tabincell{l}{Matrix computations\\Sort computations\\Sampling computations\\Transform computations\\Basic Statistic} & \tabincell{l}{Matrix construction; Matrix multiplication\\Quick sort; Min/max calculation\\Interval sampling\\FFT/IFFT Transformation\\Count Statistics} \\
  \hline
\end{tabular}
\end{table*}

Considering the mandatory requirements of simulation time and performance data accuracy for architecture community,
we implement four proxy benchmarks with respect to four representative Hadoop benchmarks from BigDataBench~\cite{wang2014bigdatabench} -- TeraSort, Kmeans, PageRank, and SIFT, according to our benchmarking methodology.
At the requests of our industry partners, we implemented the four proxy benchmarks in advance because of the following reasons. Other proxy benchmarks are being built using the same methodology.

\textbf{Representative Application Domains} They are all widely used in many important application domains. For example, TeraSort is a widely-used workload in many application domains; PageRank is a famous workload for search engine; Kmeans is a simple but useful workload used in internet services; SIFT is a fundamental workload for image feature extraction.

\textbf{Various Workload Patterns} They have different workload patterns. Hadoop TeraSort is an I/O-intensive workload; Hadoop Kmeans is a CPU-intensive workload; Hadoop PageRank is a hybrid workload which falls between CPU-intensive and I/O-intensive; Hadoop SIFT is a CPU-intensive and memory-intensive workload.

\textbf{Diverse Data Inputs} They take different data inputs. Hadoop TeraSort uses text data generated by gensort~\cite{gensort}; Hadoop Kmeans uses vector data while Hadoop PageRank uses graph data; Hadoop SIFT uses image data from ImageNet~\cite{imagenet_cvpr09}. These benchmarks are of great significance for measuring big data systems and architectures~\cite{jia_bigDataBench_subset}.

In the rest of this paper, we use Proxy TeraSort, Proxy Kmeans, Proxy PageRank and Proxy SIFT to represent the proxy benchmark for Hadoop TeraSort, Hadoop Kmeans, Hadoop PageRank and Hadoop SIFT from BigDataBench, respectively.
The input data to each proxy benchmark  has the same data type and distribution with respect to those of the Hadoop benchmarks so as to preserve the data impact on workload behaviors.
To resemble the process of the Hadoop framework, we choose the OpenMP implementations of the dwarf components, which are implemented with similar processes to the Hadoop framework, including input data partition, chunk data allocation per thread, intermediate data output to disk, and data combination.
JVM garbage collection (GC) is an important step for automatic memory management. Currently, we don't consider GC's impacts in the proxy benchmark construction, as GC only occupies a small fraction in the Hadoop workloads (about 1\% time spent on GC in our experiments) when the node has enough memory.
Finally, we use the auto-tuning tool to generate the proxy benchmarks. The process of constructing these four proxy benchmarks shows that our auto-tuning method can reach a steady state after dozens of iterations.
Table~\ref{data_workloads} lists the benchmark details from the perspectives of data set, involved dwarfs, involved dwarf components.

\section{Evaluation}\label{evaluation}


In this section, we evaluate our proxy benchmarks from the perspectives of runtime speedup and accuracy.

\subsection{Experiment Setups}

We deploy a five-node cluster, with one master node and four slave nodes. They are connected using 1Gb ethernet network. Each node is equipped with two Intel Xeon E5645 (Westmere) processors, and each processor has six physical out-of-order cores. The memory of each node is 32GB. Each node runs Linux CentOS 6.4 with the Linux kernel version 3.11.10. The JDK and Hadoop versions are 1.7.0 and 2.7.1, respectively. The GCC version is 4.8.0, which supports the OpenMP 3.1 specification.
The proxy benchmarks are compiled using "-O2" option for optimization.
The hardware and software details are listed on Table \ref{hwconfigeration}.

\begin{table}
\caption{Node Configuration Details of Xeon E5645}\label{hwconfigeration}
\renewcommand\arraystretch{1.2}
\center
\footnotesize
\begin{tabular}{|p{0.7in}|p{0.66in}|p{0.66in}|p{0.66in}|}
\hline \rowcolor{mygray} \multicolumn{4}{|l|}{Hardware Configurations}\\
\hline \multicolumn{2}{|c|}{CPU Type} & \multicolumn{2}{c|}{Intel CPU Core} \\
\hline \multicolumn{2}{|c|}{Intel \textregistered Xeon E5645}  &\multicolumn{2}{c|}{6 cores@2.40G} \\
\hline L1 DCache &L1 ICache &L2 Cache &L3 Cache \\
\hline 6 $\times$ 32 KB& 6 $\times$ 32 KB&6 $\times$ 256 KB& 12MB \\
\hline \multicolumn{2}{|c|}{Memory} & \multicolumn{2}{c|}{32GB,DDR3}  \\
\hline \multicolumn{2}{|c|}{Disk} & \multicolumn{2}{c|}{SATA@7200RPM}\\
\hline \multicolumn{2}{|c|}{Ethernet} & \multicolumn{2}{c|}{1Gb}\\
\hline \multicolumn{2}{|c|}{Hyper-Threading} & \multicolumn{2}{c|}{Disabled}\\
\hline
\hline \rowcolor{mygray} \multicolumn{4}{|l|}{Software Configurations}\\
\hline \tabincell{l}{Operating\\System} & \tabincell{l}{Linux\\Kernel} & \tabincell{l}{JDK\\Version} & \tabincell{l}{Hadoop\\Version} \\
\hline CentOS 6.4 & 3.11.10 & 1.7.0 & 2.7.1\\
\hline
\end{tabular}
\end{table}

To evaluate the performance data accuracy, we run the proxy benchmarks against the  benchmarks from BigDataBench. We run the four Hadoop benchmarks from BigDataBench on the above five-node cluster using the optimized Hadoop configurations, through tuning the data block size of the Hadoop distributed file system, memory allocation for each map/reduce job and reduce job numbers according to the cluster scale and memory size.
For Hadoop TeraSort, we choose 100 GB text data produced by gensort~\cite{gensort}. For Hadoop Kmeans and PageRank, we choose 100 GB sparse vector data with 90\% sparsity~\footnote{The sparsity of the vector indicates the proportion of zero-valued elements.} and $2^{26}$-vertex graph both generated by BDGS~\cite{ming2014bdgs}, respectively. For Hadoop SIFT, we use one hundred thousand images from ImageNet~\cite{imagenet_cvpr09}.
For comparison, we run the four proxy benchmarks on one of the slave nodes, respectively.

\subsection{Metrics Selection and Collection}

To evaluate accuracy, we choose  micro-architectural and system metrics covering instruction mix, cache behavior, branch prediction, processor performance, memory bandwidth and disk I/O behavior.
Table \ref{simulation_metric} presents the metrics we choose.

\textbf{Processor Performance}. We choose two metrics to measure the processor overall performance. Instructions per cycle (IPC) indicates the average number of instructions executed per clock cycle. Million instructions per second (MIPS) indicates the instruction execution speed.

\textbf{Instruction Mix}. We consider the instruction mix breakdown including the percentage of integer instructions, floating-point instructions, load instructions, store instructions and branch instructions.

\textbf{Branch Prediction}. Branch predication is an important strategy used in modern processors. We track the miss prediction ratio of branch instructions (br\_miss for short).

\textbf{Cache Behavior}. We evaluate cache efficiency using cache hit ratios, including L1 instruction cache, L1 data cache, L2 cache and L3 cache.

\textbf{Memory Bandwidth}. We measure the data load rate from memory and the data store rate into memory, with the unit of bytes per second. We choose metrics of memory read bandwidth (read\_bw for short), memory write bandwidth (write\_bw for short) and total memory bandwidth including both read and write (mem\_bw for short).

\textbf{Disk I/O Behavior}. We employ I/O bandwidth to reflect the I/O behaviors of workloads.


We collect micro-architectural metrics from hardware performance monitoring counters (PMCs),
and look up the hardware events' value on Intel Developer's Manual~\cite{intel2010intel}. Perf~\cite{perftool} is used to collect these hardware events. To guarantee the accuracy and validity, we run each workload three times, and collect performance data of workloads on all slave nodes during the whole runtime. We report and analyze their average value.

\begin{table}[htb]
\caption{System and Micro-architectural Metrics.}
\renewcommand\arraystretch{1.1}
\center
\footnotesize
\begin{tabular}{|p{0.68in}|p{0.86in}|p{1.25in}|}
  \hline
  Category& Metric Name &  Description \\
  \hline \rowcolor{mygray} \multicolumn{3}{|l|}{Micro-architectural Metrics}\\
  \hline
  \multirow{3}{*}{\tabincell{l}{Processor\\Performance}} & IPC & \tabincell{l}{Instructions per cycle} \\
  \cline{2-3}
   & MIPS & \tabincell{l}{Million instructions\\per second} \\
  \hline
  \multirow{4}{*}{\tabincell{l}{Instruction\\Mix}} & \multirow{4}{*}{\tabincell{l}{Instruction\\ratios}} & \multirow{4}{*}{\tabincell{l}{Ratios of floating-\\point, load, store,\\branch and integer\\instructions}}  \\
  &&\\
  &&\\
  &&\\
  \hline
  \multirow{2}{*}{\tabincell{l}{Branch\\Prediction}}  & Branch Miss & Branch miss prediction ratio \\
  \hline
  \multirow{6}{*}{\tabincell{l}{Cache\\Behavior}}  & L1I Hit Ratio & L1 instruction cache hit ratio \\
  \cline{2-3}
   & L1D Hit Ratio & L1 data cache hit ratio \\
  \cline{2-3}
   & L2 Hit Ratio & L2 cache hit ratio \\
  \cline{2-3}
   & L3 Hit Ratio & L3 cache hit ratio \\
  \hline
  \rowcolor{mygray} \multicolumn{3}{|l|}{System Metrics}\\ \hline
  \multirow{6}{*}{\tabincell{l}{Memory\\Bandwidth}}  & \tabincell{l}{Read\\Bandwidth} & \tabincell{l}{Memory load band-\\width}\\
  \cline{2-3}
   & \tabincell{l}{Write\\Bandwidth} & \tabincell{l}{Memory store band-\\width}\\
  \cline{2-3}
   & \tabincell{l}{Total\\Bandwidth} & memory load and store bandwidth \\
  \hline
  \multirow{2}{*}{\tabincell{l}{Disk I/O\\Behavior}} &  \multirow{2}{*}{\tabincell{l}{Disk I/O\\Bandwidth}} & Disk read and write bandwidth \\
  \hline
\end{tabular}\label{simulation_metric}
\end{table}

\subsection{Runtime Speedup}

Table~\ref{time-e5645} presents the execution time of the Hadoop benchmarks and the proxy benchmarks on Xeon E5645. Hadoop TeraSort with 100 GB text data runs 1500 seconds on the five-node cluster. Hadoop Kmeans with 100 GB vectors runs 5971 seconds for each iteration. Hadoop PageRank with $2^{26}$-vertex graph runs 1443 seconds for each iteration. Hadoop SIFT with one hundred thousands images runs 721 seconds.
The four corresponding proxy benchmarks run about ten seconds on the physical machine. For TeraSort, Kmeans, PageRank, SIFT, the speedup is 136X (1500/11.02), 743X (5971/8.03), 160X (1444/9.03) and 90X (721/8.02), respectively.

\begin{table}[htb]
\caption{Execution Time on Xeon E5645.}\label{time-e5645}
\renewcommand\arraystretch{1.2}
\center
\footnotesize
\begin{tabular}{|p{0.72in}|p{0.98in}|p{0.9in}|}
  \hline
  \multirow{2}{*}{Workloads} & \multicolumn{2}{c|}{Execution Time (Second)} \\
  \cline{2-3}
  & Hadoop version & Proxy version \\
  \hline
  TeraSort & 1500 & 11.02 \\
  \hline
  Kmeans & 5971 & 8.03 \\
  \hline
  PageRank & 1444 & 9.03 \\
  \hline
  SIFT & 721 & 8.02 \\
  \hline
\end{tabular}
\end{table}

\subsection{Accuracy}

We evaluate the accuracy of all metrics listed in Table \ref{simulation_metric}. For each  metric, the accuracy of the proxy benchmark comparing to the Hadoop benchmark is computed by Equation \ref{similarity-equ1}. Among which, $Val_H$ represents the average value of the Hadoop benchmark on all slave nodes; $Val_P$ represents the average value of the proxy benchmark on a slave node. The absolute value ranges from 0 to 1. The number closer to 1 indicates higher accuracy.

\begin{equation} \label{similarity-equ1}
\begin{aligned}
Accuracy(Val_H,Val_P)=1-\left|\frac{Val_P-Val_H}{Val_H}\right|
\end{aligned}
\end{equation}

\begin{figure}[htb]
\centering
\includegraphics*[scale=0.6]{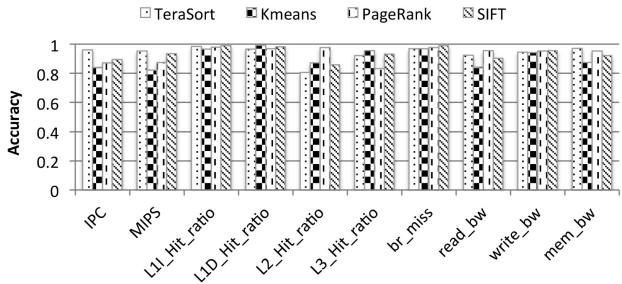}
\caption{System and Micro-architectural Data Accuracy on Xeon E5645.}
\label{e5645:metric}
\end{figure}

\begin{figure}[htb]
\centering
\includegraphics*[scale=0.54]{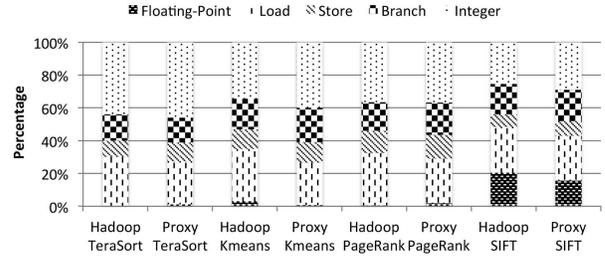}
\caption{Instruction Mix Breakdown on Xeon E5645.}
\label{e5645:inst}
\end{figure}

Fig.~\ref{e5645:metric} presents the system and micro-architectural data accuracy of the proxy benchmarks on Xeon E5645. We can find that the average accuracy of all metrics are greater than 90\%. For TeraSort, Kmeans, PageRank, SIFT, the average accuracy is 94\%, 91\%, 93\%, 94\%, respectively.
Fig.~\ref{e5645:inst} shows the instruction mix breakdown of the proxy benchmarks and Hadoop benchmarks.
From Fig.~\ref{e5645:inst}, we can find that the four proxy benchmarks preserve the instruction mix characteristics of these four Hadoop benchmarks. For example, the integer instruction occupies 44\% for Hadoop TeraSort and 46\% for Proxy TeraSort, while the floating-point instruction occupies less than 1\% for both Hadoop and Proxy TeraSort. For instructions involving data movement, Hadoop TeraSort contains 39\% of load and store instructions, and Proxy TeraSort contains 37\%. The SIFT workload, widely used in computer vision for image processing, has many floating-point instructions. Also,  Proxy SIFT preserves the instruction mix characteristics of Hadoop SIFT.

\subsubsection{Disk I/O Behaviors}

Big data applications have significant disk I/O pressures. To evaluate the DISK I/O behaviors of the proxy benchmarks, we compute the disk I/O bandwidth using Equation \ref{io-bandwidth}, where $Sector_{Read+Write}$ means the total number of sector reads and sector writes; $Size_{Sector}$ means the sector size (512 bytes for our nodes).

\begin{equation} \label{io-bandwidth}
\begin{aligned}
BW_{Disk I/O}=\frac{(Sector_{Read+Write})*Size_{Sector}}{RunTime}
\end{aligned}
\end{equation}

\begin{figure}[htb]
\centering
\includegraphics*[scale=0.6]{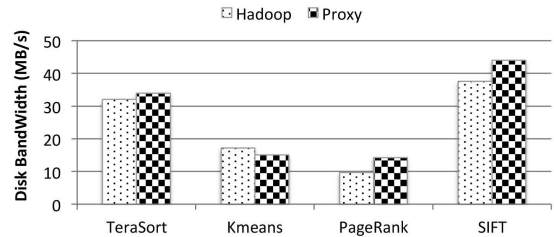}
\caption{Disk I/O BandWidth on Xeon E5645.}
\label{e5645:io}
\end{figure}

Fig.~\ref{e5645:io} presents the I/O bandwidth of proxy benchmarks and Hadoop benchmarks on Xeon E5645. We can find that they have similar average disk I/O pressure. The disk I/O bandwidth of Proxy TeraSort and Hadoop TeraSort is 32.04 MB and 33.99 MB per second, respectively.

\subsubsection{The Impact of Input Data}

\begin{figure}[htb]
\centering
\includegraphics*[scale=0.7]{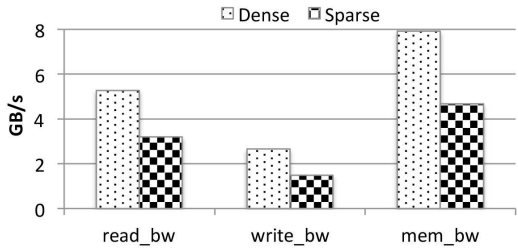}
\caption{Data Impact on Memory Bandwidth Using Sparse and Dense Data for Hadoop Kmeans on Xeon E5645.}
\label{e5645:sparse:diff}
\end{figure}

In this section, we demonstrate when we change the input data sparsity, our proxy benchmarks can still mimic the big data workloads with a high accuracy.
We run  Hadoop Kmeans with the same Hadoop configurations, and the data input is 100 GB.  
We use different data input: sparse vector (the original configuration, 90\% elements are zero) and dense vectors (all elements are non-zero, and 0\% elements are zero). 
Fig.~\ref{e5645:sparse:diff} presents the performance difference using different data input.
 We can find that the memory bandwidth with sparse vectors is nearly half of the memory bandwidth with dense vectors, which confirms the data input's impacts on micro\comp architectural performance.

\begin{figure}[htb]
\centering
\includegraphics*[scale=0.68]{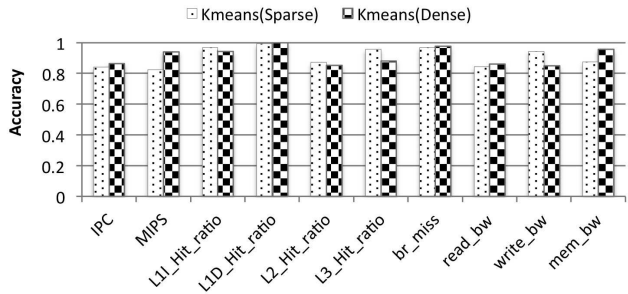}
\caption{System and Micro-architectural Data Accuracy Using Different Data Input on Xeon E5645.}
\label{e5645:dense:metric}
\end{figure}

On the other hand, Fig. \ref{e5645:dense:metric}  presents the accuracy of proxy benchmark using diverse input data.
We can find that  the average micro\comp architectural data accuracy of Proxy Kmeans is above 91\%  with respect to the fully-distributed Hadoop Kmeans using dense input data with no zero-valued element. When we change the input data sparsity from 0\% to 90\%, the data accuracy of Proxy Kmeans is also above 91\% with respect to the original workload. So we see that the Proxy Kmeans can mimic the Hadoop Kmeans  under different input data.


\section{Case Studies on ARM Processors}

This section demonstrates our proxy benchmark also can mimic the original benchmarks on the ARMv8 processors.
We report our joint evaluation work with our industry partnership on ARMv8 processors using Hadoop TeraSort, Kmeans, PageRank, and the corresponding proxy benchmarks. Our evaluation includes widely acceptable metrics: runtime speedup, performance accuracy and several other concerns like multi-core scalability and cross-platform speedup.

\subsection{Experiment Setup}

Due to the resource limitation of ARMv8 processors, we use a two-node (one master and one slave) cluster with each node equipped with one ARMv8 processor. In addition, we deploy a two-node (one master and one slave) cluster with each node equipped with one Xeon E5-2690 v3 (Haswell) processor for speedup comparison.
Each ARMv8 processor has 32 physical cores, with each core having independent L1 instruction cache and L1 data cache. Every four cores share L2 cache and all cores share the last-level cache. The memory of each node is 64GB.
Each Haswell processor has 12 physical cores, with each core having independent L1 and L2 cache. All cores share the last-level cache. The memory of each node is 64GB. In order to narrow the gap of logical core numbers between two architectures, we enable hyperthreading for Haswell processor. Table \ref{config} lists the hardware and software details of two platforms.

Considering the memory size of the cluster, we use 50 GB text data generated by gensort for Hadoop TeraSort, 50 GB dense vectors for Hadoop Kmeans, and $2^{24}$-vertex graph data for Hadoop PageRank.
We run Hadoop benchmarks with optimized configurations, thr-\\ough tuning the data block size of the Hadoop distributed file system, memory allocation for each job and reduce task numbers according to the cluster scale and memory size.
 For comparison, we run proxy benchmarks on the slave node.
 Our industry partnership pays great attentions on cache and memory access patterns, which are important micro-architectural and system metrics for chip design. So we mainly collect cache-related and memory-related performance data.

\begin{table}
\caption{Platform Configurations} { \scriptsize 
\begin{tabular}{|p{0.84in}|p{0.82in}|p{1.15in}|}
\hline \rowcolor{mygray} \multicolumn{3}{|l|}{Hardware Configurations}\\
\hline Model&ARMv8 & Xeon E5-2690 V3\\
\hline Number of Processors & 1 & 1\\
\hline Number of Cores  & 32 & 12\\
\hline Frequency&2.1GHz&2.6GHz\\
\hline L1 Cache(I/D)&48KB/32KB&32KB/32KB\\
\hline L2 Cache&8 x 1024KB&12 x 256KB\\
\hline L3 Cache&32MB&30MB\\
\hline Architecture&ARM&X86\_64\\
\hline Memory&64GB, DDR4&64GB, DDR4\\
\hline Ethernet&1Gb&1Gb\\
\hline Hyper-Threading & None & Enabled\\
\hline
\hline \rowcolor{mygray} \multicolumn{3}{|l|}{Software Configurations}\\
\hline \tabincell{l}{Operating\\System} & \tabincell{l}{EulerOS V2.0} &\tabincell{l}{Red-hat Enterprise Linux\\Server release 7.0}\\
\hline Linux Kernel & 4.1.23-vhulk3.6.3.aarch64 & 3.10.0-123.e17.x86-64\\
\hline GCC Version & 4.9.3 & 4.8.2\\
\hline JDK Version & jdk1.8.0\_101 & jdk1.7.0\_79\\
\hline \tabincell{l}{Hadoop Version} & 2.5.2 & 2.5.2\\
\hline\end{tabular} } \label{config}
\end{table}\

\subsection{Runtime Speedup on ARMv8}

Table.~\ref{time-arm} presents the execution time of Hadoop benchmarks and the proxy benchmarks on ARMv8. Our proxy benchmarks run within 10 seconds on ARMv8 processor.
On two-node cluster equipped with ARMv8 processor, Hadoop TeraSort with 50 GB text data runs 1378 seconds. Hadoop Kmeans with 50GB vectors runs 3374 seconds for one iteration. Hadoop PageRank with $2^{24}$-vertex runs 4291 seconds for five iterations. In contrast, their proxy benchmarks  run 4102, 8677 and 6219 milliseconds, respectively.
For TeraSort, Kmeans, PageRank, the speedup is 336X (1378/4.10), 386X (3347/8.68) and 690X (4291/6.22), respectively.

\begin{table}[htb]
\caption{Execution Time on ARMv8.}\label{time-arm}
\renewcommand\arraystretch{1.1}
\center
\footnotesize
\begin{tabular}{|p{0.72in}|p{0.98in}|p{0.9in}|}
  \hline
  \multirow{2}{*}{Workloads} & \multicolumn{2}{c|}{Execution Time (Second)} \\
  \cline{2-3}
  & Hadoop version & Proxy version \\
  \hline
  TeraSort & 1378 & 4.10 \\
  \hline
  Kmeans & 3347 & 8.68 \\
  \hline
  PageRank & 4291 & 6.22 \\
  \hline
\end{tabular}
\end{table}

\subsection{Accuracy on ARMv8}

\begin{figure}[htb]
\centering
\includegraphics*[scale=0.6]{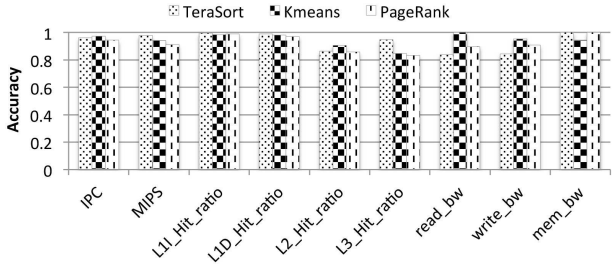}
\caption{System and Micro-architectural Data Accuracy on ARMv8.}
\label{arm:metric}
\end{figure}

We report the system and micro-architectural data accuracy of the Hadoop benchmarks and the proxy benchmarks. Likewise, we evaluate the accuracy by Equation \ref{similarity-equ1}.
Fig.~\ref{arm:metric} presents the accuracy of the proxy benchmarks on ARM processor. We can find that on the ARMv8 processor, the average data accuracy is all above 90\%.
For TeraSort, Kmeans and PageRank, the average accuracy is 93\%, 95\% and 92\%, respectively.

\subsection{Multi-core Scalability on ARMv8}

ARMv8 has 32 physical cores, and we evaluate its multi-core scalability using the Hadoop benchmarks and the proxy benchmarks on 4, 8, 16, 32 cores, respectively. For each experiment, we disable the specified number of cpu cores through cpu-hotplug mechanism. For the Hadoop benchmarks, we adjust the Hadoop configurations so as to get the peak performance. For the proxy benchmarks, we run them directly without any modification.

\begin{figure*}[htbp]
\centering
\subfloat[TeraSort]{
\label{core:sort}
\begin{minipage}[t]{.33\linewidth}
\centering
\includegraphics[scale=0.47]{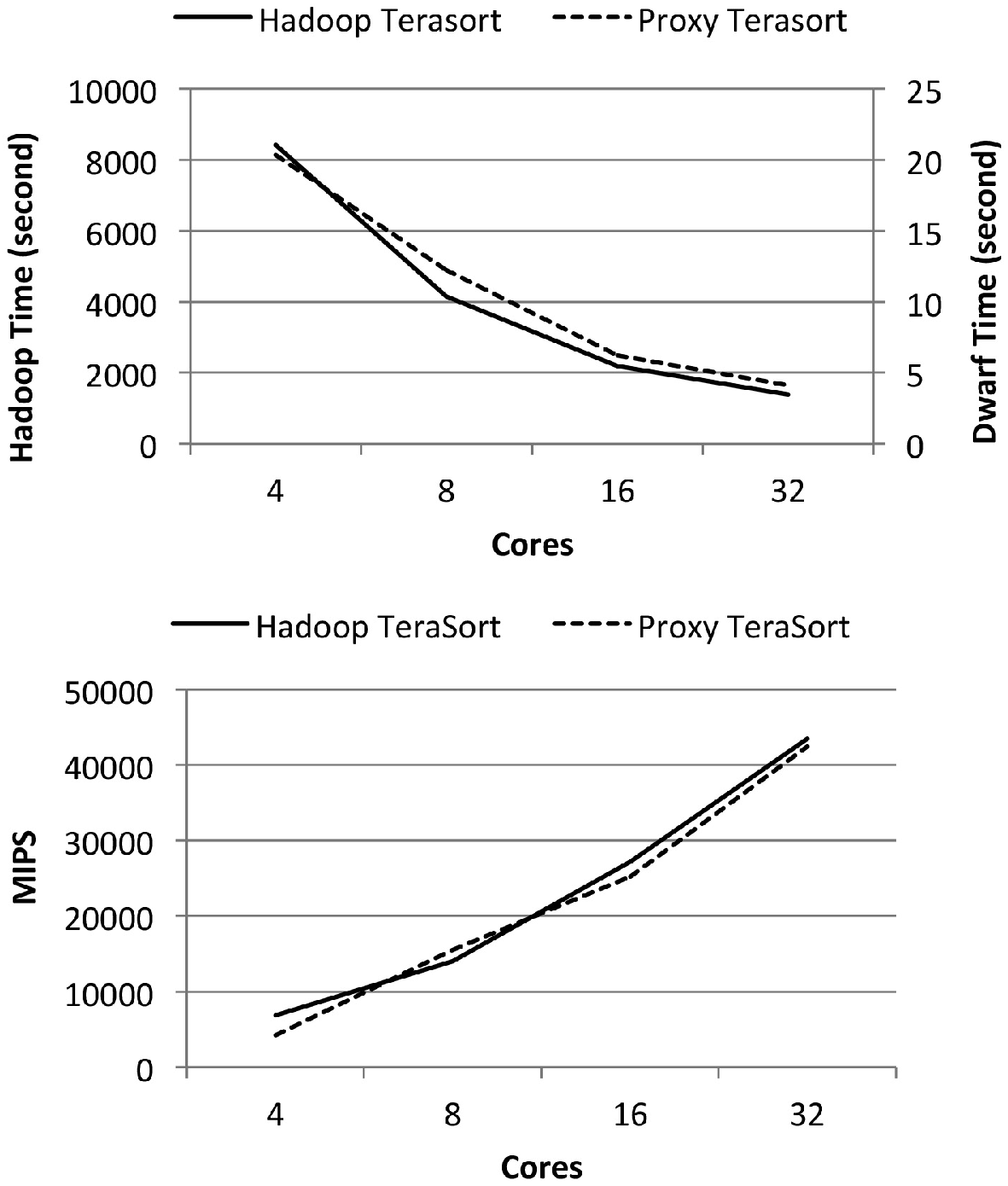}
\end{minipage}
}
\subfloat[Kmeans]{
\label{core:kmeans}
\begin{minipage}[t]{.33\linewidth}
\centering
\includegraphics[scale=0.47]{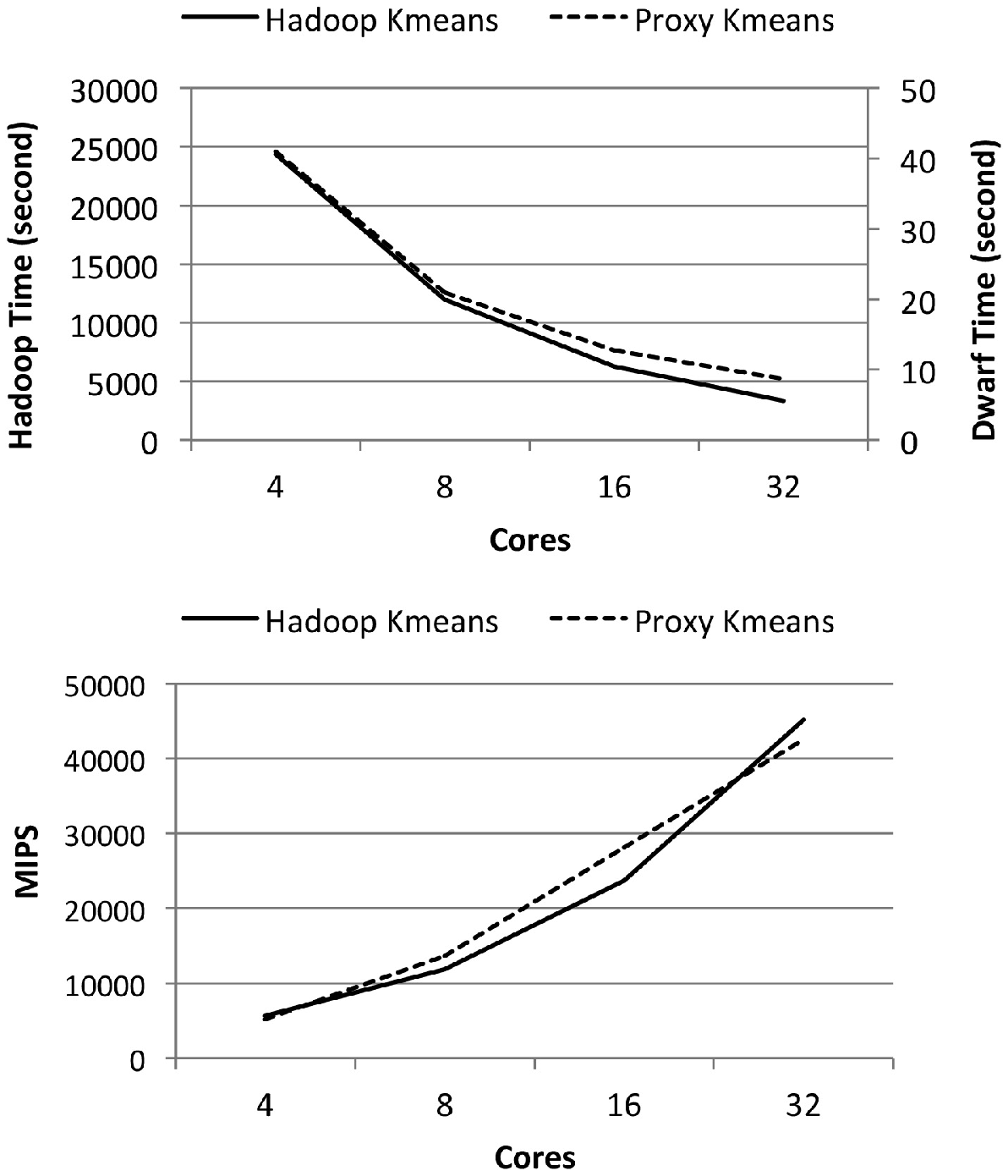}
\end{minipage}
}
\subfloat[PageRank]{
\label{core:pagerank}
\begin{minipage}[t]{.33\linewidth}
\centering
\includegraphics[scale=0.47]{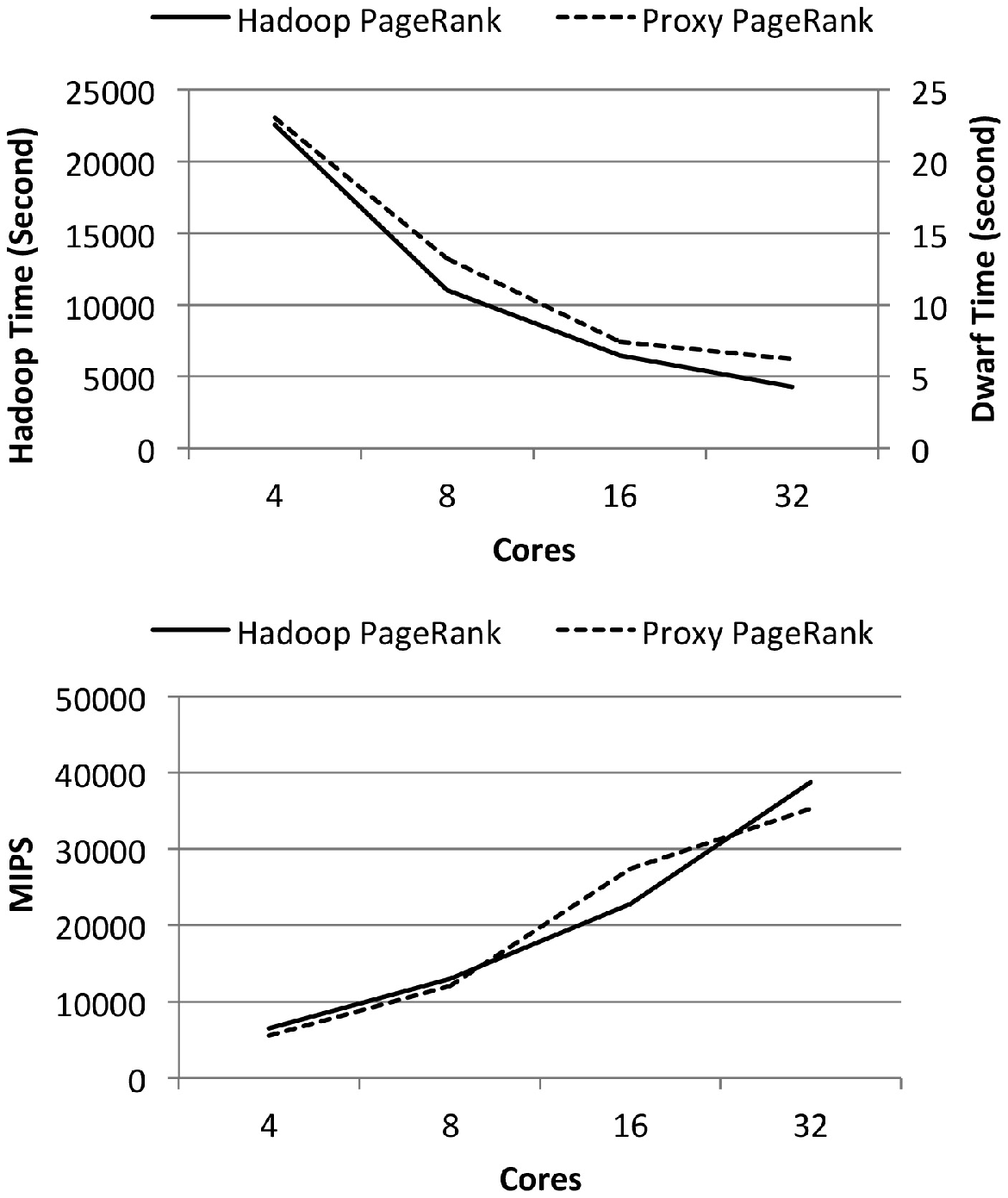}
\end{minipage}
}
\caption{Multi-core Scalability of the Hadoop benchmarks and Proxy Benchmarks on ARMv8.}
\label{core:all}
\end{figure*}

Fig.~\ref{core:all} reports multi-core scalability in terms of runtime and MIPS. The horizontal axis represents the core number and the vertical axis represents runtime or MIPS. Due to the large runtime gap between the Hadoop benchmarks and proxy benchmarks, we list their runtime on different side of vertical axis: the left side indicates runtime of the Hadoop benchmarks, while the right side indicates runtime of the proxy benchmarks. We can find that they have similar multi-core scalability trends in terms of both runtime and instruction execution speed.

\subsection{Runtime Speedup across Different Processors}

Runtime speedup across different processors is another metric with much concern from our industry partnership.
In order to reflect the impacts of different design decisions for running big data analytics workloads on X86 and ARM architectures, the proxy benchmarks should be able to maintain consistent performance trends.
That is to say, if the proxy benchmarks gain performance promotion through an improved design, the real workloads can also benefit from the design.
We evaluate the runtime speedup across two different architectures of ARMv8 and Xeon E5-2690 V3 (Haswell).
The runtime speedup is computed using Equation \ref{speedup-time}. The Hadoop configurations are also optimized according to hardware environments. The proxy benchmarks use the same version on two architectures.

\begin{equation} \label{speedup-time}
\begin{aligned}
Speedup(Time_{X86\_64},Time_{ARM})=\frac{Time_{ARM}}{Time_{X86\_64}}
\end{aligned}
\end{equation}

\begin{figure}[!t]
\centering
\includegraphics[scale=0.6]{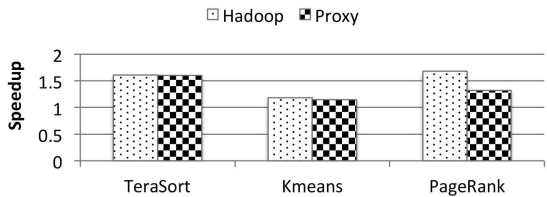}
\caption{Runtime Speedup across Different Processors.} 
\label{speedup}
\end{figure}

Fig.~\ref{speedup} shows the runtime speedups of the Hadoop benchmarks and the proxy benchmarks across ARM and X86\_64 architectures. We can find that they have consistent speedup trends. For example, Hadoop TeraSort runs 1378 seconds and 856 seconds on ARMv8 and Haswell, respectively. Proxy TeraSort runs 4.1 seconds and 2.56 seconds on ARMv8 and Haswell, respectively. The runtime speedups between ARMv8 and Haswell are 1.61 (1378/856) running Hadoop TeraSort, and 1.60 (4.1/2.56) running Proxy TeraSort.

\section{Related Work}

Multiple benchmarking methodologies have been proposed over the past few decades.
The most simplest one is to create a new benchmark for every possible workload.
PARSEC~\cite{bienia11benchmarking} provides a series of shared-memory programs for chip-multiprocessors.  BigDataBench~\cite{wang2014bigdatabench} is a benchmark suite providing dozens of big data workloads. CloudSuite~\cite{ferdman2011clearing} consists of eight applications, which are selected based on popularity. These benchmarking methods need to provide individual implementations for every possible workload, and keep expanding benchmark set to cover  emerging workloads. Moreover, it is frustrating to run (component or application) benchmarks like BigDataBench or CloudSuite on simulators because of complex software stacks and long running time.
Using reduced data input is one way to reduce execution time. Previous work~\cite{kleinosowski2001adapting,kleinosowski2002minnespec} adopts reduced data set for the SPEC benchmark and maintains similar architecture behaviors using the full reference data sets. 

Kernel benchmarks are widely used in  high performance computing. Livermore kernels~\cite{mcmahon1986livermore} use Fortran applications to measure floating-point performance range.
 The NAS parallel benchmarks~\cite{bailey1991parallel} consist of several separate tests, including five kernels and three pseudo-applications derived from computational fluid dynamics (CFD) applications.
Linpack~\cite{dongarra2003linpack} provides a collection of Fortran subroutines. Kernel benchmarks are insufficient to completely reflect workload behaviors considering the complexity and diversity of big data workloads\cite{bailey1991parallel,lilja2005measuring}.

In terms of micro-architectural simulation, many previous studies generated synthetic benchmarks as proxies~\cite{bell2005improved,ganesan2010synthesizing}. 
Statistical simulation~\cite{skadron2003challenges,eeckhout2004control,eeckhout2000performance,nussbaum2001modeling,oskin2000hls,eeckhout2001early} generates synthetic trace or synthetic benchmarks to mimic micro-architectural performance of long-running real workloads, which targets one workload on a specific architecture with the certain configurations, and thus each benchmark needs to be generated on the other architectures with different  configurations~\cite{joshi2007constructing}.
Sampled simulation selects a series of sample units for simulation instead of entire instruction stream, which were sampled randomly~\cite{conte1996reducing}, periodically~\cite{wunderlich2003smarts,yu2009tss} or based on phase behavior~\cite{sherwood2002automatically}.
Seongbeom et al.~\cite{kim2007accelerating} accelerated the full-system simulation through characterizing and predicting the performance behavior of OS services.
For emerging big data workloads, PerfProx~\cite{panda2017proxy} proposed a proxy benchmark generation framework for real-world database applications through characterizing low-level dynamic execution characteristics.

Our big data dwarfs are  inspired by previous successful abstractions in other application scenarios.
The \emph{set} concept in relational algebra~\cite{codd1970relational} abstracted five primitive and fundamental operators (Select, Project, Product, Union, Difference), setting off a wave of relational database research. The set abstraction is the basis of relational algebra and theoretical foundation of database.
Phil Colella~\cite{colella2004defining} identified seven dwarfs of numerical methods which he thought would be important for the next decade. Based on that, a multidisciplinary group of Berkeley researchers proposed 13 dwarfs which were highly abstractions of parallel computing, capturing the computation and communication patterns of a great mass of applications~\cite{asanovic2006landscape}.
National Research Council proposed seven major tasks in massive data analysis~\cite{council2013frontiers}, which they called giants. These seven giants are macroscopical definition of problems from the perspective of mathematics, while our eight classes of dwarfs are frequently-appearing units of computation in the above tasks and problems.

\section{Conclusions}

In this paper, we propose a dwarf-based scalable big data benchmarking methodology.
After thoroughly analyzing a majority of algorithms in five typical application domains: search engine, social networks, e-commerce, multimedia, and bioinformatics, we capture eight big data dwarfs among a wide variety of big data analtyics workloads, including matrix, sampling, logic, transform, set, graph, sort and basic statistic computation.
For each dwarf, we implement the dwarf components  using diverse software stacks. We construct and tune the big data proxy benchmarks using the DAG-like combinations of dwarf components using different wights to mimic  the  benchmarks in BigDataBench.

Our proxy benchmarks can reach 100s runtime speedup with respect to the benchmarks from BigDataBench, while the average micro-architectural data accuracy is above 90\% on both X86\_64 and ARM architectures. The proxy benchmarks have been applied to ARM processor design in our industry partnership.

\bibliographystyle{ieeetr}

\end{document}